\documentclass{jps-cp}
\usepackage{txfonts} 
\usepackage{graphicx}
\usepackage{bm}
\usepackage{color}

\newcommand{\mathsym}[1]{{}}
\newcommand{\unicode}[1]{{}}
\newcommand{\be}{\begin{eqnarray}}
\newcommand{\ee}{\end{eqnarray}}

\title{Phenomenology of the chiral $d$-wave state in the hexagonal pnictide superconductor SrPtAs}

\author{Hikaru \textsc{Ueki}$^{1}$, Shoma \textsc{Inagaki}$^{1}$, Ryota \textsc{Tamura}$^{1}$, Jun \textsc{Goryo}$^{1}$, 
Yoshiki \textsc{Imai}$^{2}$, W. B. \textsc{Rui}$^{3}$, Andreas P. \textsc{Schnyder}$^{3}$, and  Manfred \textsc{Sigrist}$^{4}$}

\inst{$^{1}$Department of Mathematics and Physics, Hirosaki University, Hirosaki 036-8561, Japan \\
$^{2}$Department of Applied Physics., Okayama University of Science, 1-1 Ridaicho, Kita-Ku, 700-0005 Okayama, Japan \\
$^{3}$Max-Planck-Institute f\"{u}r Festk\"{o}rperforschung, Heisenbergstrasse 1, D-70569 Stuttgart, Germany \\
$^{4}$Institute f\"{u}r Theoretische Physik, ETH Z\"{u}rich, 8093 Z\"{u}rich, Switzerland }

\email{hikaruueki@hirosaki-u.ac.jp}

\recdate{September 15, 2019}

\abst{The pairing symmetry of the hexagonal pnictide superconductor SrPtAs is discussed 
with taking into account its multiband structure. 
The topological chiral $d$-wave state with time-reversal-symmetry breaking has been anticipated from the spontaneous magnetization observed by the muon-spin-relaxation experiment.  We point out in this paper that the recent experimental reports on the nuclear-spin-lattice relaxation rate $T_1^{-1}$ and superfluid density $n_s(T)$, which seemingly support the conventional $s$-wave pairing, are also consistent with the chiral $d$-wave state. The compatibility of the gap and multiband structures is crucial in this argument. 
We propose that the measurement of the bulk quasiparticle density of states would be useful for the distinction between two pairing states.}

\kword{Chiral superconductivity, Topological superconductivity, Multiband superconductivity, Nuclear-spin-lattice relaxation rate, Superfluid density}

\begin{document}
\maketitle

\section{Introduction}
The first hexagonal pnictide superconductor SrPtAs \cite{Nishikubo-Kudo-Nohara} ($T_c=2.4 K$) has attracted attention,
since the internal  spontaneous magnetization is observed by
the muon-spin-relaxation ($\mu$SR) experiment below $T_c$ \cite{Biswas-etal}. 
As the result, the spontaneous time-reversal symmetry (TRS) breaking in
the superconducting state is suggested. 
The most probable pairing symmetry 
suggested by the group theoretical consideration \cite{Goryo-Fischer-Sigrist,Fischer-Goryo} 
and functional renormalization group (FRG) analysis \cite{Fischer-etal}
is the topological chiral $d$-wave ($d_{x^2-y^2} \pm i d_{xy}$-wave) state with TRS breaking. 
This state has non-zero Chern number\cite{TKNN} and supports the surface bound states with chiral energy spectrum\cite{Volovik,Fischer-etal}. 
Especially in SrPtAs, it is expected that 
the chiral surface state causes spontaneous spin current and spin polarization \cite{Goryo-etal-spin}, 
origin of which is the staggered anti-symmetric spin-orbit coupling (SOC) coming from
the hexagonal bi-layer structure of the crystal with local lack of inversion symmetry \cite{Sigrist-review}.

On the other hand, there are still some controversies on the chiral $d$-wave state of SrPtAs. 
The nuclear spin-lattice relaxation rate $T_1^{-1}$ measured by the nuclear quadrupole resonance 
shows the Hebel-Slichter (HS) peak near $T_c$ and exponential decay in the low temperature region \cite{Matano-etal}.  
It has also been found from the magnetic-penetration-depth measurement that superfluid density $n_s(T)$ 
exhibits the Arrhenius-type behavior (i.e., approaches to $n_s(0)$ exponentially) 
at low temperature\cite{Landaeta-etal}. 
The conventional $s$-wave pairing without any nodal excitation is naively expected from these experimental results. 
Moreover, the other FRG analysis for this superconductor suggestes the $f$-wave pairing, 
although it takes into account the contribution from the dominant band \cite{f-wave}.

Using the multiband quasiclassical theory \cite{Nagai-etal},  
we address this issue in this paper and show that observed $T_1^{-1}$ and $n_s(T)$ are consistent with 
the chiral $d$-wave pairing as well as the $s$-wave one. 
We also show that the experimental observation of the bulk quasiparticle DOS, 
which can be measured by the scanning tunneling spectroscopy/microscopy (STM/STS), 
would be crucial for the distinction between $s$- and chiral $d$-wave pairing states.

\begin{figure}
\begin{center}
\centering
\includegraphics[width=0.4\linewidth]{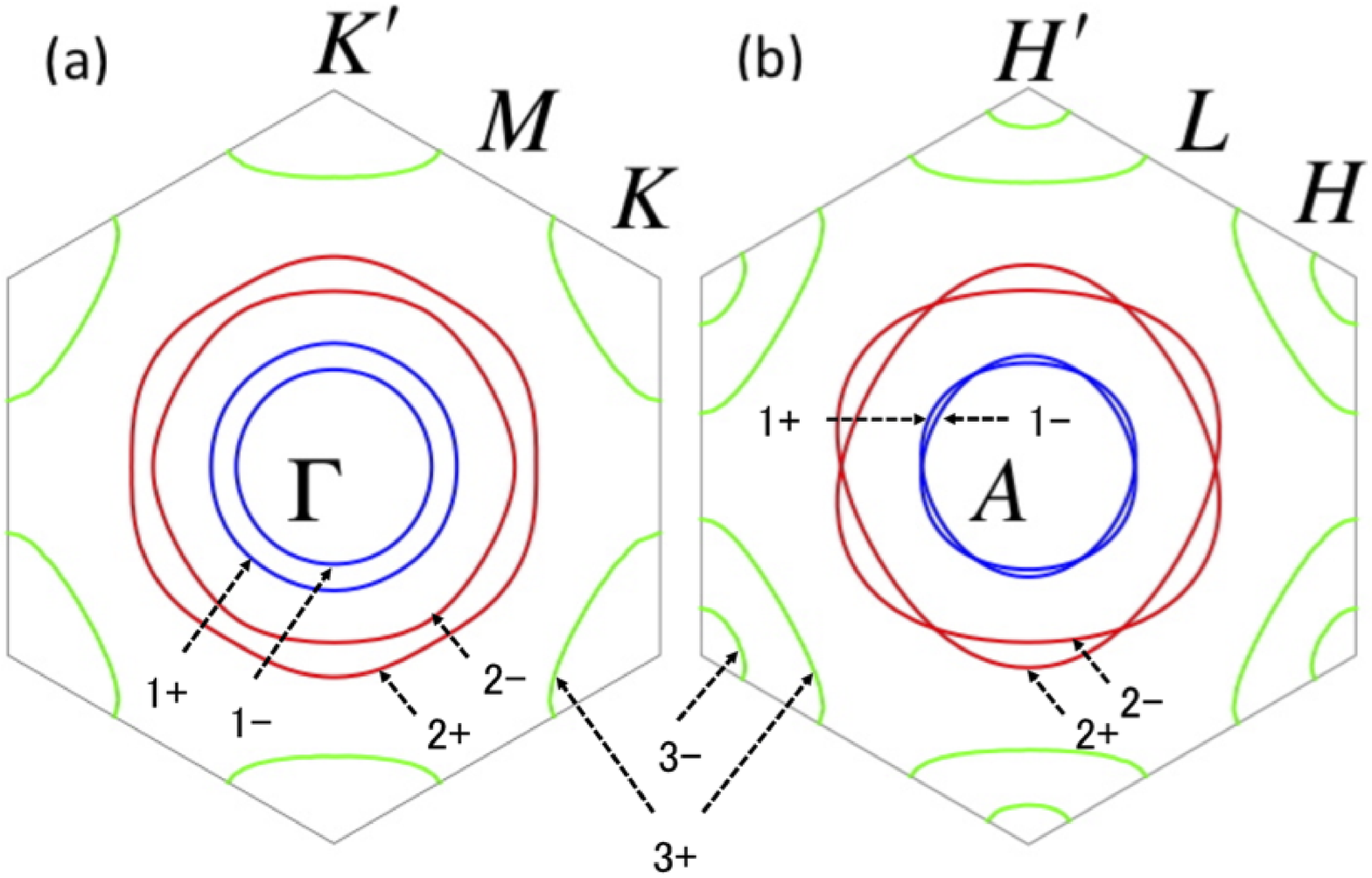}
\end{center} 
\caption{The cross section of Fermi surfaces at (a) $k_z=0$ and (b) $k_z=\pi / c$. 
Each Fermi surface is labelled by the set of parameters $\beta=1,2,3$ and $\gamma=\pm$ (see also Eq. (\ref{normal_spectrum})).}
\label{FS}
\end{figure}

\section{Normal and pairing states}
The normal-state energy spectrum of SrPtAs is \cite{Youn-etal,Youn-etal-2}
\begin{eqnarray}
\xi^{\beta \gamma}_{\bm k}=\epsilon^{(\beta)}_{1\bm k}-\mu^{(\beta)} + \gamma \sqrt{|\epsilon^{(\beta)}_{c \bm k}|^2+|\alpha^{(\beta)} \lambda_{\bm k}|^2} 
\label{normal_spectrum}
\end{eqnarray}
where $\beta(=1,2,3)$ indicates the unsplit band, 
$\gamma=\pm$ is the band splitting parameter, 
$a_{\bm k l \sigma}^{(\beta)\dagger}$ and $a_{\bm k l \sigma}^{(\beta)}$ are 
the creation and annihilation operators of an electron 
with the wave vector $\bm k$ and spin $\sigma=\pm 1$ in the $l$-th layer of the unit cell, 
$\epsilon^{(\beta)}_{1 \bm k}=t^{(\beta)}_{1} \sum_n \cos {\bm k}\cdot {\bm T}_n+t^{(\beta)}_{c2}\cos (k_z c)$, 
$\epsilon^{(\beta)}_{c \bm k}=t^{(\beta)}_{c} \cos (k_z c/2)[1+\exp(-i\bm k \cdot \bm T_3)+\exp(i\bm k \cdot \bm T_2)]$, 
and $\lambda_{\bm k}=\sum_n \sin \bm k \cdot {\bm T}_n$ 
with $\bm T_1=(0,a,0)$, $\bm T_2=(\sqrt{3}a/2,-a/2,0)$, and $\bm T_3=(-\sqrt{3}a/2,-a/2,0)$ the 
in-plane nearest-neighbor bond vectors ($a$ and $c$ are in-plane and inter-layer lattice constants).
We use the tight-binding parameters suggested by the LDA calculation \cite{Youn-etal,Youn-etal-2}, 
and obtain the Fermi surface structure depicted as Fig. \ref{FS}.

We adopt the quasiclassical formalism for the multiband superconductor \cite{Nagai-etal}. 
The Green's functions of the $``\beta\gamma"$-th bands are 
\be
g_{\uparrow\uparrow}(i \epsilon_n, \bm k_F^{\beta\gamma})=\bar{g}_{\downarrow\downarrow}(i \epsilon_n, \bm k_F^{\beta\gamma})=\frac{\epsilon_n}{\sqrt{\epsilon_n^2+|\Delta_{\bm k_F^{\beta\gamma}}|^2}}, 
\ \ \ 
f_{\uparrow\downarrow}(i \epsilon_n, \bm k_F^{\beta\gamma})=\left\{\bar{f}_{\downarrow\uparrow}(i \epsilon_n, \bm k_F^{\beta\gamma})\right\}^*=
\frac{\Delta_{\bm k_F^{\beta\gamma}}}{\sqrt{\epsilon_n^2+|\Delta_{\bm k_F^{\beta\gamma}}}|^2},   
\nonumber
\ee
where $\epsilon_n=(2n+1) \pi k_B T$ is the fermionic Matsubara energy, 
$\bm k_F^{\beta\gamma} $ denotes the Fermi wave vector,
and $\Delta_{\bm k_F^{\beta\gamma}}$ denotes the gap function. 
We simply assume the gap function as \cite{Nagai-etal,note-eq4}
\be
\Delta_{\bm k_F^{\beta\gamma}}=\Delta(T) \phi_{\bm k_F^{\beta\gamma}}, \ \ \
\Delta(T)=
\left\{
\begin{array}{cc}
\Delta_0 \tanh \left[\frac{\pi k_B T_c}{\Delta_0}\sqrt{\delta\left(\frac{T_c}{T}-1\right)}\right] & (T\leq T_c)
\\
0 & (T>T_c)
\end{array}
\right., \label{gap}
\ee
with $\delta=1.05$, and $\phi_{\bm k_F^{\beta\gamma}}$ shown in Table \ref{phi-k} is 
the long-wavelength expansion (around the center of $``\beta\gamma"$ th Fermi surface)  
of the tight-binding pair wave functions for $s$-, $f$-, and chiral $d$-wave states, 
$1$, $\sum_{n=1}^3 \sin \bm k \cdot \bm T_n$, 
and $\sum_{n=1}^3 e^{i 2 \pi n /3} \cos \bm k \cdot \bm T_n$, respectively \cite{Goryo-Fischer-Sigrist,Fischer-Goryo}.

Let us introduce phenomenologically the quasiparticle damping (the smearing factor of the quasiparticle DOS) $\eta$ 
via the analytic continuation to obtain the retarded and advanced Green's functions 
\be
&g_{\uparrow\uparrow}^{R,A}(\epsilon, \bm k_F^{\beta\gamma})=g_{\uparrow\uparrow}(i\epsilon_n \rightarrow \epsilon \pm i \eta, \bm k_F^{\beta\gamma}), \ \ \ 
\bar{g}_{\downarrow\downarrow}^{R,A}(\epsilon, \bm k_F^{\beta\gamma})=\bar{g}_{\downarrow\downarrow}(i\epsilon_n \rightarrow \epsilon \pm i \eta, \bm k_F^{\beta\gamma}), \nonumber\\
&f_{\uparrow\downarrow}^{R,A}(\epsilon, \bm k_F^{\beta\gamma})=f_{\uparrow\downarrow}(i\epsilon_n \rightarrow \epsilon \pm i \eta, \bm k_F^{\beta\gamma}), \ \ \
\bar{f}_{\downarrow\uparrow}^{R,A}(\epsilon, \bm k_F^{\beta\gamma})=\bar{f}_{\downarrow\uparrow}(i\epsilon_n \rightarrow \epsilon \pm i \eta, \bm k_F^{\beta\gamma}). \label{Green-RA}
\ee
Here we have neglected the band dependence of $\eta$. 
Thus, there are two fitting parameters, $\Delta_0/k_B T_c$ and $\eta$, in the following calculations.

\begin{table}
\caption{The list of $\phi_{{\bm k}_F^{\beta\gamma}}$, which is the long-wavelength expansions (around the center of $``\beta\gamma"$ th Fermi surface) for $s$-, $f$-, and chiral $d$-wave pair wave functions 
in the tight-binding scheme \cite{Goryo-Fischer-Sigrist, Fischer-Goryo}. 
We obtain the normalization constant from the condition 
$\langle |\phi_{{\bm k}_F}^{\beta\gamma}|^2 \rangle_{F^{\beta\gamma}}=1$, 
where $\langle \cdot\cdot\cdot \rangle_{F^{\beta\gamma}}$ denotes the average on the $\beta\gamma$ th Fermi surface.
The abbreviations $3\pm(H)$ th and $3\pm(H')$ th mean 
the disconnected Fermi pockets of $3\pm$ th band enclosing $H$ and $H'$ points, respectively. 
Note that all the Fermi surfaces are quasi-2D, except for the 3D $3-$ th one. 
Here, $\hat{\bm k}=\bm k/|\bm k|$, and 
$\delta{\bm k}=\bm k_F^{\beta\gamma}-\bm k_0$,
$\delta{\bm p}=\bm k_F^{\beta\gamma}-\bm p_0$,
$\delta{\bm p}'=\bm k_F^{\beta\gamma}-\bm p'_0$,
$\delta{\bm q}=\bm k_F^{\beta\gamma}-\bm q_0$,
and 
$\delta{\bm q}'=\bm k_F^{\beta\gamma}-\bm q'_0$ 
refer to the deviations from the centers of the long-wavelength expansions, 
and 
$\bm  k_0=(0,0,k_{F z}^{\beta\gamma})$,
$\bm  p_0=(2\pi/\sqrt{3},2\pi/3,k_{F z}^{\beta\gamma})$,
$\bm  p'_0=(0,4\pi/3,k_{F z}^{\beta\gamma})$,
$\bm  q_0=(2\pi/\sqrt{3},2\pi/3,\pi/c)$,
and 
$\bm  q'_0=(0,4\pi/3,\pi/c)$ the centers of the expansions.}
\begin{tabular}{c|cccccc}\hline\hline
& $1\pm$ th
& $2\pm$ th
& $3+(H)$ th
& $3+(H')$ th 
& $3-(H)$ th 
& $3-(H')$ th 
\\\hline
$\phi_{{\bm k}_F^{\beta\gamma}}$ of $s$-wave
& $1$
& $1$
& $1$
& $1$
& $1$
& $1$
\\
$\phi_{{\bm k}_F^{\beta\gamma}}$ of $f$-wave
& $(3 \delta \hat{k}_x^2-\delta \hat{k}_y^2) \delta \hat{k}_y$
& $(3 \delta \hat{k}_x^2-\delta \hat{k}_y^2) \delta \hat{k}_y$
& $1$
& $-1$
& $1$
& $-1$
\\
$\phi_{{\bm k}_F^{\beta\gamma}}$ of chiral $d$-wave
& $(\delta\hat{k}_{x} + i \delta\hat{k}_{y})^2$
& $(\delta\hat{k}_{x} + i \delta\hat{k}_{y})^2$
& $\delta\hat{p}_{x} - i \delta\hat{p}_{y}$
& $\delta \hat{p}'_{x} - i \delta\hat{p}'_{y}$
& $\delta\hat{q}_{x} - i \delta\hat{q}_{y}$
& $\delta \hat{q}'_{x} - i \delta\hat{q}'_{y}$
\\\hline\hline
\end{tabular}
\label{phi-k}
\end{table}

\section{The nuclear-spin-lattice relaxation rate $T_1^{-1}$} 
The relaxation rate is given by \cite{Nagai-etal} 
\be
\frac{T_1(T_c)}{T_1(T)}=\frac{T}{T_c} \int_{-\infty}^{\infty} d \epsilon \left(\bar{N}_{s}(\epsilon)^2+\bar{M}_{s}(\epsilon)^2\right)\left(-\frac{\partial f(\epsilon)}{\partial \epsilon}\right), 
\label{1/T1}
\ee
where $f(\epsilon)$ denotes the Fermi-Dirac distribution function, and 
$\bar{N}_s(\epsilon)$ and $\bar{M}_s(\epsilon)$ denote DOS and anomalous DOS of the Bogoliubov quasiparticle normalized by the bulk normal DOS at the Fermi level $N(0)$. 
For the multiband spin-singlet superconductor, 
$\bar{N}_s(\epsilon)$ and $\bar{M}_s(\epsilon)$ are \cite{Nagai-etal}
\be
\left\{
\begin{array}{l}
\bar{N}^2_{s}(\epsilon)=\langle a_{\uparrow\uparrow}^{11}(-\epsilon, \bm k) \rangle_F \langle a_{\downarrow\downarrow}^{22}(\epsilon, \bm k) \rangle_F, 
\\
\bar{M}^2_{s}(\epsilon)=-\langle a_{\uparrow\downarrow}^{12}(-\epsilon, \bm k) \rangle_F \langle a_{\downarrow\uparrow}^{21}(\epsilon, \bm k) \rangle_F,
\end{array}
\right.
\ee
where 
\be
&a_{\uparrow\uparrow}^{11}(\epsilon, \bm k) = \frac{1}{2} \left(g_{\uparrow\uparrow}^R(\epsilon, \bm k)-g_{\uparrow\uparrow}^A(\epsilon, \bm k)\right), \ \ \ 
a_{\downarrow\downarrow}^{22}(\epsilon, \bm k) = \frac{1}{2}\left(\bar{g}_{\downarrow\downarrow}^R(\epsilon, \bm k)-\bar{g}_{\downarrow\downarrow}^A(\epsilon, \bm k)\right),
\nonumber\\  
&a_{\uparrow\downarrow}^{12}(\epsilon, \bm k) = \frac{i}{2} \left(f_{\uparrow\downarrow}^R(\epsilon, \bm k)- f_{\uparrow\downarrow}^A(\epsilon, \bm k)\right), \ \ \
a_{\downarrow\uparrow}^{21}(\epsilon, \bm k) = \frac{i}{2} \left(\bar{f}_{\downarrow\uparrow}^R(\epsilon, \bm k)- \bar{f}_{\downarrow\uparrow}^A(\epsilon, \bm k)\right),  
\ee 
and $\langle \cdots \rangle_F$ is the Fermi surface average
\be
\langle a_{\sigma\sigma'}^{\tau\tau'}(\epsilon, \bm k) \rangle_F = \frac{1}{N(0)}\sum_{\beta\gamma} \int \frac{d \Omega_{\bm k_{F}^{\beta\gamma}}}{(2 \pi)^3 \hbar |\bm v_{F}^{\beta\gamma}|} a_{\sigma\sigma'}^{\tau\tau'}(\epsilon, \bm k_F^{\beta\gamma}),
\ee
with the Fermi velocity $\hbar \bm v_F^{\beta\gamma}=\left.{\bm \nabla}_{\bm k} \xi_{\bm k}^{\beta\gamma}\right|_{\bm k=\bm k_F^{\beta\gamma}}$.

\begin{figure}
\begin{minipage}{0.47\hsize}
\begin{center}
\centering
\includegraphics[width=0.9\linewidth]{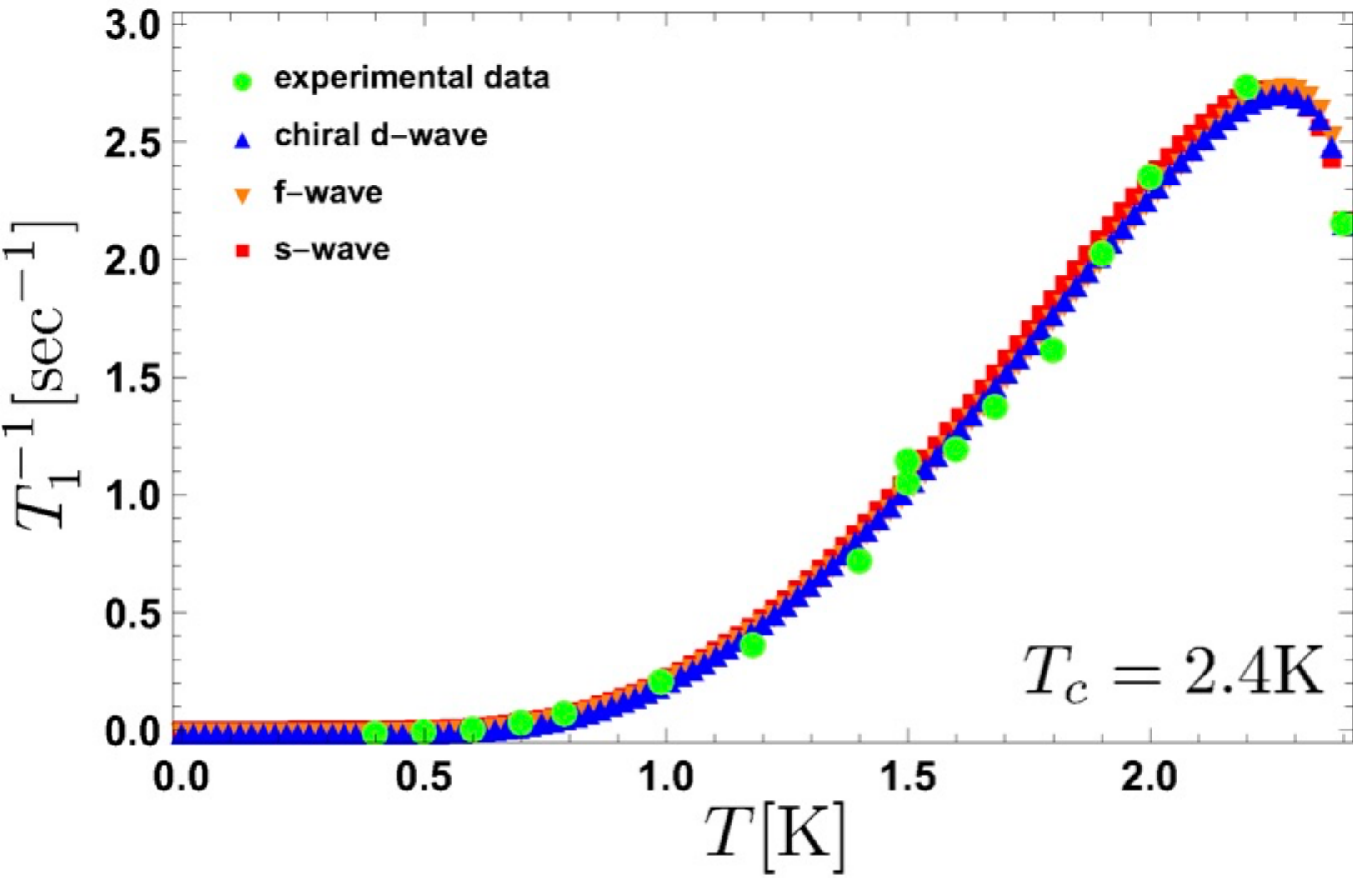}
\end{center} 
\caption{Temperature dependence of $T_1^{-1}$. Green dots are the experimental results \cite{Matano-etal}. 
Red squares, blue triangles, and orange triangles 
show the estimations for $s$-wave, chiral $d$-wave, and $f$-wave states. 
}
\label{T1}
\end{minipage}
\ \ \ \ \ \ \ \ 
\begin{minipage}{0.47\hsize}
\begin{center}
\centering
\includegraphics[width=0.9\linewidth]{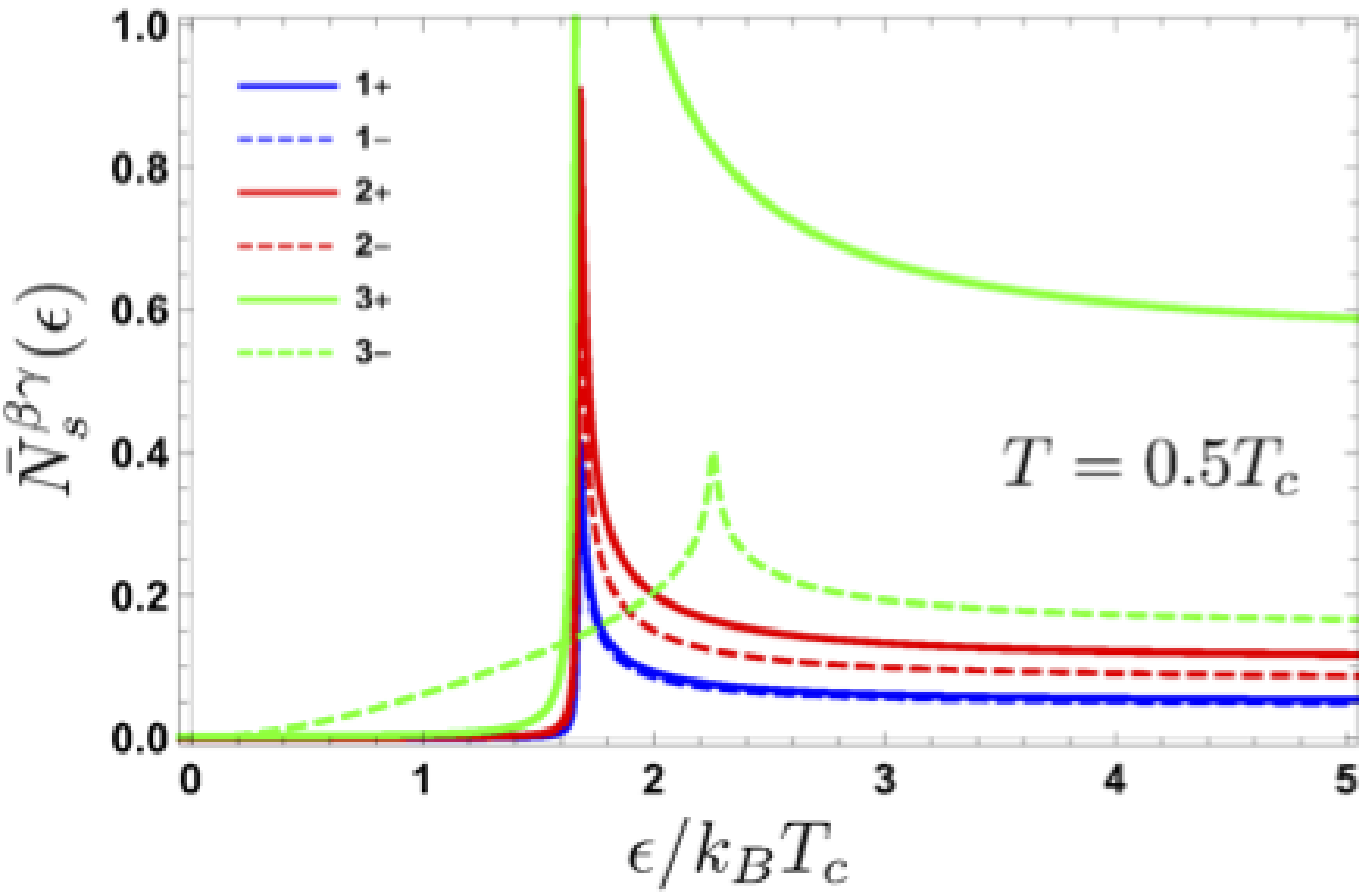}
\end{center} 
\caption{The contrubution from the ``$\beta\gamma$"-th band to the normalized quasiparticle DOS $\bar{N}_s(\epsilon)$ in the chiral $d$-wave state, which is expressed as 
$\bar{N}_s^{\beta\gamma}(\epsilon)= \int a_{\uparrow\uparrow}^{11}(\epsilon, \bm k_F^{\beta\gamma}) d \Omega_{\bm k_{F}}^{\beta\gamma} /\left\{(2 \pi)^3 N(0) \hbar |\bm v_{F}^{\beta\gamma}|\right\}$. 
The gap amplitude at $T=0.5T_c$ is used in this estimation.  
 }
\label{SC_DOS-each}
\end{minipage}
\end{figure}

The results for the three pairing states are shown in Fig. \ref{T1} 
with experimental data\cite{Matano-etal}. 
The fitting parameters are chosen as $\Delta_0/k_B T_c=1.765$ for all states, 
and $\eta=0.14 k_B T_c$ for the $s$-wave state, 
$\eta=0.0025 k_B T_c$ for the $f$-wave state, 
and $\eta=0.008 k_B T_c$ for the chiral $d$-wave state.
We clearly see that experimental data can be fitted well by all pairing states
showing the HS peak just below $T_c$ and exponential decay at low temperature.

It is renown that $\bar{M}_s(\epsilon)$ from the coherence effect appears only for the $s$-wave pairing 
and contributes to the HS peak significantly \cite{Thinkham-textbook}. 
However, the quasiparticle excitations of quasi-2D bands  (1 $\pm$ th and 2 $\pm$ th) 
in the chiral $d$-wave state are fully gapped 
and also give rise to the large enough HS peak by reduced $\eta$ in $\bar{N}_s(\epsilon)$.
Moreover, the bulk quasiparticle DOS of the 3D (3 $-$ th) band is less dominant in this system 
as seen in Fig. \ref{SC_DOS-each}, 
and then the power-low behavior near zero temperature caused by the nodal excitation is negligible. 
These facts are crucial for the compatibility of the chiral $d$-wave pairing with experiment data. 
On the other hand, 
the quasiparticle excitation in the $f$-wave state is fully gapped in two bands (3 $\pm$ th) around the Brillouin zone corners, 
whereas has line nodes in four quasi-2D bands (1 $\pm$ th and 2 $\pm$ th) 
around the zone center. 
The result for the $f$-wave state can also fit well with observed data \cite{Matano-etal} with reduced $\eta$, 
thanks to the fully-gapped 3 $+$ th band. 
We also note additionally that the gap maxima, 
which determines the peak location, of the point nodal gap function $\Delta(T) \phi_{{\bm k}_F^{3-}}$ is larger than 
that of the other nodeless gap functions due to the normalization condition 
$\langle |\phi_{{\bm k}_F}^{\beta\gamma}|^2 \rangle_{F^{\beta\gamma}}=1$ [see also Eq. (\ref{gap})].

\section{Superfluid density $n_s (T)$} 
Superfluid density normalized by its zero-temperature value $n_{s}(0)$ is \cite{Nagai-etal} 
\be
\bar{n}_s(T)=\sum_{\beta\gamma} \bar{n}_s^{\beta\gamma}(T), \ \ \ \ \
\bar{n}_s^{\beta\gamma}(T)=\frac
{\sum_{i=x,y,z} \int \frac{d \Omega_{\bm k_{F}^{\beta\gamma}}}{(2 \pi)^3 \hbar |\bm v_{F}^{\beta\gamma}|}   \left(v_{Fi}^{\beta\gamma}\right)^2 \left(1 - Y_{\bm k_F^{\beta\gamma}}(T)\right)}
{\sum_{\beta\gamma} \sum_{i=x,y,z} \int \frac{d \Omega_{\bm k_{F}^{\beta\gamma}}}{(2 \pi)^3 \hbar |\bm v_{F}^{\beta\gamma}|}   \left(v_{Fi}^{\beta\gamma}\right)^2}, \label{sup-dens}
\ee
where $Y_{\bm k}(T)$ is Yosida function
\be
Y_{\bm k}(T)=1-\pi k_B T \sum_{n=-\infty}^{\infty} \frac{|\Delta_{\bm k}|^2}{\left(\epsilon_n^2 + |\Delta_{\bm k}|^2\right)^{3/2}}.
\ee
The parameter is taken as $\Delta_0/k_B T_c=1.5$ for all states \cite{note}. 
We have also checked that the results are insensitive to the choice of the smearing factor $\eta$ in this case, 
and we choose the same values of $\eta$ used in the calculations of $T_1^{-1}$.

\begin{figure}
\begin{minipage}{0.47\hsize}
\begin{center}
\centering
\includegraphics[width=0.9\linewidth]{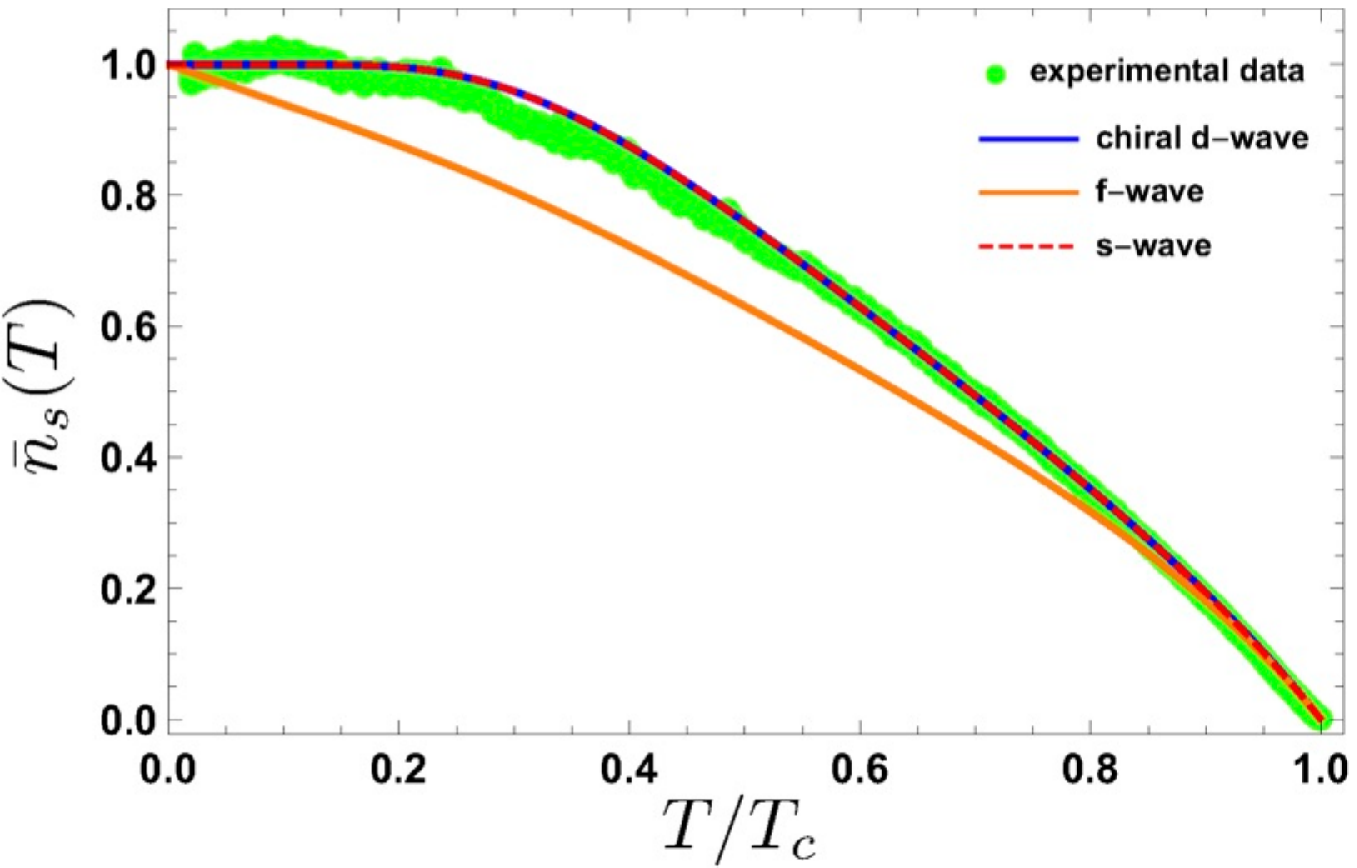}
\end{center} 
\caption{Temperature dependence of normalized superfluid density $\bar{n}_s(T)$. Green dots denote experimental data \cite{Landaeta-etal}.  
Dashed red, blue, and orange lines show the estimations 
for $s$-wave, chiral $d$-wave, and $f$-wave states. 
}
\label{n_s}
\end{minipage}
\ \ \ \ \ \ \ \ 
\begin{minipage}{0.47\hsize}
\begin{center}
\centering
\includegraphics[width=0.9\linewidth]{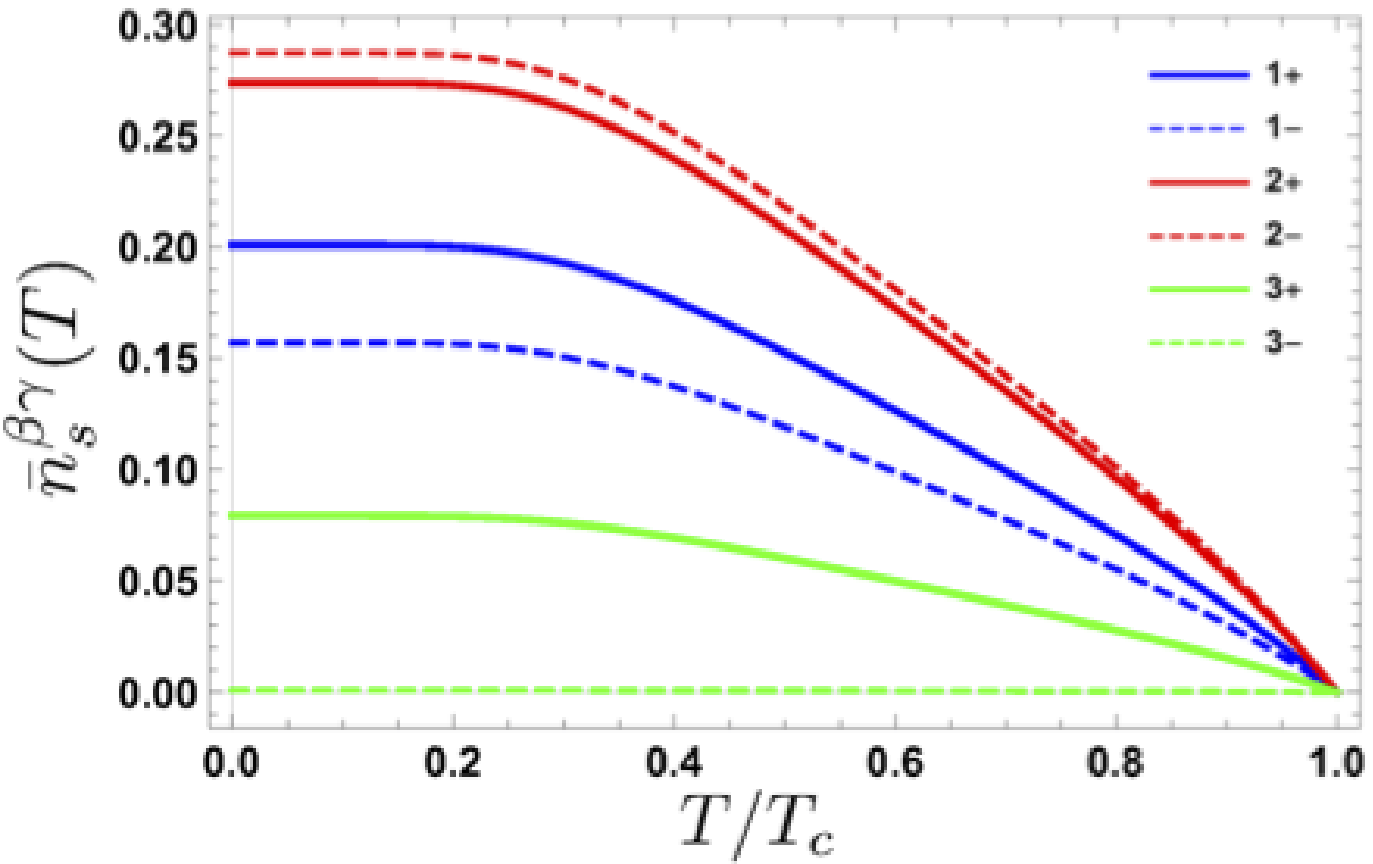}
\end{center} 
\caption{$\bar{n}_s^{\beta\gamma}(T)$ of the chiral $d$-wave state. 
We see that the contribution from the 3D ($3-$ th) band with power-law behavior is negligibly small.}
\label{FS-dep-n_s}
\end{minipage}
\end{figure}

We clearly see from Fig. \ref{n_s} that the results of $s$- and chiral $d$-wave states fit very well to experimental data, namely, both exhibit the thermal-activation-type behavior at low temperature. 
$\bar{n}_s^{\beta\gamma}(T)$ in Eq. (\ref{sup-dens}) shows the contribution from each band and the result for the chiral $d$-wave state is plotted in Fig. \ref{FS-dep-n_s}. 
The contribution to $n_s(T)$ from the 3D band 
with power-law behavior is negligibly small, since $\bar{n}_s^{\beta\gamma}(T)$
depends strongly on root mean square of the Fermi velocity, and its value for the 3D band is minor (see Table. I in Ref. \cite{Youn-etal-2}). 
On the other hand, $n_s(T)$ in the $f$-wave state shows an evident power-law behavior at low temperature 
and contradicts strongly with the experiment \cite{Landaeta-etal}. 
The power-law behavior is due to the large root mean square of the Fermi velocity of the line nodal $1 \pm$ th and $2 \pm$ th bands.

\begin{figure}
\begin{center}
\centering
\includegraphics[width=0.4\linewidth]{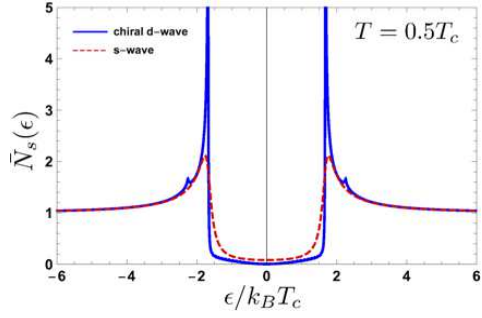}
\end{center} 
\caption{Dashed red and blue lines are the normalized DOS of quasiparticles $\bar{N}_s(\epsilon)$ in $s$- and chiral $d$-wave states with the gap amplitude at $T=0.5T_c$. 
The smearing factor $\eta=0.14 (0.008) k_B T_c$, and the peaks are reduced (enhanced) in the $s$-wave (chiral $d$-wave) state. }
\label{SC_DOS}
\end{figure}

We therefore cannot distinguish between $s$- and chiral $d$-wave states from $T_1^{-1}$ and $n_s (T)$. 
{\it We then propose that the experimental observation of the bulk quasiparticle DOS would give a decisive distinction.} 
We need to reduce $\eta$ for the chiral $d$-wave state to compensate the absence of the contribution from $\bar{M}_s(\epsilon)$. Namely,   
the reduction of $\eta$ causes the significant difference of $\bar{N}_s(\epsilon)$ for $s$- and chiral $d$-wave states 
as seen in Fig. \ref{SC_DOS}. 
Another possibility for the distinction using quasiparticle interference spectroscopy has also been pointed out 
\cite{Akbari-Thalmeier}.

\section{Summary} 
We have shown based on the multiband quasiclassical theory \cite{Nagai-etal} that 
$T_1^{-1}$ and $n_s(T)$ observed in the superconducting phase of SrPtAs \cite{Matano-etal,Landaeta-etal} 
are consistent with the chiral $d$-wave state as well as the $s$-wave one. 
On the other hand, $f$-wave state is inconsistent with $n_s(T)$. 
We have found in the fitting of $T_1^{-1}$ a significant difference of 
the quasiparticle damping factors for $s$- and chiral $d$-wave pairing states due to 
the absence of $\bar{M}_s(\epsilon)$ in the chiral $d$-wave state as seen in Eq. (\ref{1/T1}). 
This difference causes a remarkable difference between the magnitudes of the peaks in the bulk quasiparticle DOS, 
therefore, a measurement of which would 
give a decisive distinction between $s$- and chiral $d$-wave states as seen in Fig. \ref{SC_DOS}. 
The bulk quasiparticle DOS could be observed by the STM/STS 
even in the (0001) surface without the chiral surface mode. 
It should be emphasized that the chiral $d$-wave state is the only state which is compatible with all the experiments 
that have been done so far \cite{Biswas-etal,Matano-etal,Landaeta-etal}.

\section*{Acknowledgments }
The authors thank M. Nohara and K. Kudo for their useful discussions, and I. Bonalde and K. Matano for sending their experimental data. 
This work was partially supported by JSPS KAKENHI Grant Number 15H05885 (J-Physics).  
J.G. is grateful to the Pauli Center for Theoretical Physics of ETH Zurich for hospitality.


\begin{thebibliography}{9}
\bibitem{Nishikubo-Kudo-Nohara} Y. Nishikubo, K. Kudo, and M. Nohara, J. Phys. Soc. Jpn. {\bf 80}, 055002 (2011).
\bibitem{Biswas-etal} P. K. Biswas, H. Luetkens, T. Neupert, T. St\"{u}rzer, C. Baines, G. Pascua, A. P. Schnyder, M. H. Fischer, J. Goryo, M. R. Lees, H. Maeter, F. Br\"{u}ckner, H.-H. Klauss, M. Nicklas, P. J. Baker, A. D. Hillier, M. Sigrist, A. Amato, and D. Johrendt, Phys. Rev. B {\bf 87}, 180503(R) (2013). 
\bibitem{Goryo-Fischer-Sigrist} J. Goryo, M. H. Fischer, and M. Sigrist, Phys. Rev. B {\bf 86}, 100507(R) (2012). 
\bibitem{Fischer-Goryo} M. H. Fischer, and J. Goryo, J. Phys. Soc. Jpn. {\bf 84}, 054705 (2015); {\it ibid.} {\bf 86},
068001 (2017). 
\bibitem{Fischer-etal} M. H. Fischer, T. Neupert, C. Platt, A. P. Schnyder, W. Hanke, J. Goryo, R. Thomale, and M. Sigrist, Phys. Rev. B {\bf 89}, 020509(R) (2014);  {\it ibid.} {\bf 90}, 099902 (2014). 
\bibitem{TKNN} D. J. Thouless, M. Kohmoto, M. P. Nightingale, and M. den Nijs, Phys. Rev. Lett. {\bf 49}, 405 (1982).
\bibitem{Volovik} G. E. Volovik, JETP Letters {\bf 66}, 522 (1997). 
\bibitem{Goryo-etal-spin} J. Goryo, Y. Imai, W. B. Rui, M. Sigrist, and A. P. Schnyder, Phys. Rev. B {\bf 96}, 140502(R) (2017). 
\bibitem{Sigrist-review} M. Sigrist, D. F. Agterberg, M. H. Fischer, J. Goryo, F. Loder, S.-H. Rhim, D. Maruyama, Y. Yanase, T. Yoshida, and S. J. Youn, J. Phys. Soc. Jpn. {\bf 83}, 061014 (2014).   
\bibitem{Youn-etal} S. J. Youn, M. H. Fischer, S. H. Rhim, M. Sigrist, and D. F. Agterberg, Phys. Rev. B {\bf 85} 220505 (2012).
\bibitem{Youn-etal-2}S. J. Youn, S. H. Rhim, D. F. Agterberg, M. Weinert, A. J. Freeman, arXiv1202.1604.
\bibitem{f-wave} W.-S. Wang, Y. Yang, and Q.-H. Wang, Phys. Rev. B {\bf 90}, 094514 (2014).  
\bibitem{Matano-etal} K. Matano, K. Arima, S. Maeda, Y. Nishikubo, K. Kudo, M. Nohara, and Guo-qing Zheng, 
Phys. Rev. B {\bf 89}, 140504(R) (2014).  
\bibitem{Landaeta-etal} J. F. Landaeta, S. V. Taylor, I. Bonalde, C. Rojas, Y. Nishikubo, K. Kudo, and M. Nohara, 
Phys. Rev. B {\bf 93}, 064504 (2016).  
\bibitem{Nagai-etal} Y. Nagai, N. Hayashi, N. Nakai, H. Nakamura, M. Okumura, and M. Machida, New J. Phys. {\bf 10}, 103026 (2008). 
\bibitem{Thinkham-textbook}  M. Thinkham, {\it ``Introduction to Superconductivity"}  2nd ed. (Dover Publications, Inc., 1996).
\bibitem{note-eq4} 
Neglecting the band dependence of the gap amplitude $\Delta(T)$ is actually an oversimplification, but it would not be crucial. 
We would obtain essentially the same results by introducing the band dependence of the quasiparticle damping 
(see $\eta$ in Eq. (\ref{Green-RA})), even if we take band-dependent gap amplitudes into account.
\bibitem{Akbari-Thalmeier} A. Akbari and P. Thalmeier, Europhys. Lett. {\bf 106}, 27006 (2014). 
\bibitem{note} We have taken different values of the fitting parameter $\Delta_0/k_B T_c$ for $T_1^{-1}$ and $n_s(T)$. 
It seems to be natural, since different samples are used in these experiments.\cite{Matano-etal,Landaeta-etal}
\end{thebibliography}
\end{document}